\documentclass[12pt,letterpaper, twoside]{article}

\usepackage{amsmath,amsfonts,amsthm,amssymb}

\usepackage{graphicx}

\newcommand{\unit}[1]{\ensuremath{\, \mathrm{#1}}}

\usepackage{fancyhdr}
\renewcommand{\title}[1]{\noindent\textbf{\LARGE #1 } \\} 
\renewcommand{\author}[1]{\noindent\textbf{\large #1 } \\}
\renewcommand{\abstract}[1]{\noindent\textbf{\textbf{Abstract\\}} #1 \\}
\providecommand{\keywords}[1]{\noindent\textbf{\textbf{Keywords\\}} #1 \\}

\makeatletter
\let\@fnsymbol\@arabic%
\makeatother

\makeatletter
\renewcommand\@makefntext[1]{\leftskip=2em\hskip-2em\@makefnmark#1}
\makeatother

\makeatletter
\def\blfootnote{\xdef\@thefnmark{}\@footnotetext}
\makeatother

\renewcommand{\footnoterule}{%
  \kern -3pt
  \hrule width \textwidth
  \kern 2pt
}

\usepackage[labelfont=bf, justification=raggedright]{caption}
\makeatletter
\renewcommand\@biblabel[1]{#1.}
\makeatother

\begin{document}
\vspace*{-0.75in}
\noindent{\it Article}

\vspace{-5pt}

\noindent \makebox[\textwidth]{\hrulefill} \\

\title{A damage model for fracking}

\author{J Quinn Norris\textsuperscript{1}, Donald L Turcotte\textsuperscript{2} and John B Rundle\textsuperscript{1,2,3}}
\date{}

\abstract{Injections of large volumes of water into tight shale reservoirs allows the extraction of oil and gas not previously accessible. This large volume ``super" fracking induces damage that allows the oil and/or gas to flow to an extraction well. The purpose of this paper is to provide a model for understanding super fracking. We assume that water is injected from a small spherical cavity into a homogeneous elastic medium. The high pressure of the injected water generates hoop stresses that reactivate natural fractures in the tight shales. These fractures migrate outward as water is added creating a spherical shell of damaged rock. The porosity associated with these fractures is equal to the water volume injected. We obtain an analytic expression for this volume. We apply our model to a typical tight shale reservoir and show that the predicted water volumes are in good agreement with the volumes used in super fracking.}

\keywords{Damage, hydraulic fracturing, fracture, fracking, induced permeability}

\blfootnote{\textsuperscript{1}Department of Physics, University of California-Davis, United States}
\blfootnote{\textsuperscript{2}Department of Earth and Planetary Sciences, University of California-Davis, United States}
\blfootnote{\textsuperscript{3}Santa Fe Institute, United States}

\blfootnote{}
\blfootnote{\bf Corresponding author:}
\blfootnote{J Quinn Norris, Department of Physics, University of California-Davis, United States}
\blfootnote{Email: jqnorris@ucdavis.edu}
\thispagestyle{plain}

\section*{Introduction}
Injections of large volumes of water into tight shale reservoirs allows the extraction of oil and gas not previously accessible. The large volume injections were made possible by the use of ``slickwater'' beginning in the 1990's. Slickwater includes additives that reduce the water's viscosity by an order of magnitude. This reduces the resistance to flow. This large volume ``super'' fracking induces distributed damage that allows oil and/or gas to flow to an extraction well.

In order to understand super fracking it is necessary to understand the history and structure of tight shale reservoirs. During deposition, the depth and temperature of the shale increases. The increased temperature first converts the carbon to oil and subsequently to gas. This thermally activated conversion generates high fluid pressures that generate natural hydrofractures which allows some of the oil and gas to escape the reservoir and reduce the pressure. However, a large fraction of the oil and/or gas remain in the reservoir.

Subsequently, the natural hydrofractures are sealed by the deposition of silica or carbonates. This sealing leads to a ``tight'' (low permeability) reservoir. The injection of slickwater during a super frack opens the preexisting sealed fractures allowing the oil and/or gas to migrate to the production well.

The purpose of this paper is to present an idealized model for the damage generated in a super frack. We assume water is injected from a small spherical cavity in a homogeneous elastic medium. The high pressure of the injected slickwater generates hoop stresses that reactivate the sealed natural fractures in the tight shales. These fractures migrate outward as slickwater is added creating a spherical shell of damaged rock. The porosity generated is equal to the water volume injected. We obtain an analytic expression for the volume. We apply our model to a typical tight shale reservoir and show that the predicted volumes are in good agreement with the volumes of slickwater used in a super frack.

Shales that are source rocks for hydrocarbons are known as black shales due to their color. Black shales contain some 2-20\% porosity filled with organic material. Typical grain sizes are less than $4 \unit{\mu m}$, and capillary forces strongly restrict granular flows of fluids. With increasing burial depth the increasing temperature first produces oil from the organic material (the oil window) and at higher temperatures the oil breaks down to produce gas (the gas window).

The generation of oil and gas in black shales increases the fluid pressure resulting in extensional hydraulic fractures (natural fracking). Secor \cite{Secor1965} described these fractures as extensional fractures perpendicular to the least principle stress direction. A consequence of this natural fracking is the joint (fracture) sets that are found in all black shales in which oil and gas have been generated \cite{Olson2009}. Engelder et al. \cite{Engelder2009} have carried out extensive studies of the joint sets and find that they tend to be planar, parallel, and quasi-periodic with spacings in the range $0.1-0.3 \unit{m}$.

Natural fracking provides fracture permeability in shales. Granular permeability, although very low, allows oil and gas to flow to the closely spaced joints. The joint sets provide pathways for the vertical migration of the oil and gas. This oil migration has two consequences: (1) The oil flows upward into reservoirs of high permeability strata (often porous sandstones) overlain by very low permeability strata that trap the oil and gas. A large fraction of traditional oil and gas production has been from these reservoirs that are relatively easy to access. (2) A second consequence of the vertical migration is surface hydrocarbon seeps.

A tight-shale formation is defined to be a shale in which the natural fractures do not yet exist, or have been sealed, often by pressure solution and deposition of silica or carbonates. The fracture permeability is very low. In order to extract oil and gas from tight-shale formations, super fracking was developed. Fracking or hydraulic fracturing is the high pressure injection of water to create one or more open fractures in the target reservoir. A perforation is made in the well casing and high pressure water is pumped at high pressure through the perforation. The objective is to create hydraulic fractures through which the oil and gas can migrate to production wells \cite{Fjaer2008, Yew1997}.

It is important to distinguish between two types of fracking. Traditional (low volume) fracking typically uses $75-300 \unit{m^3}$ of water. Guar gum or hydroxyethyl cellulose is added to increase the viscosity of the water. The objective is to create a single or, at most, a few large fractures through which oil and gas will flow to the production well. A large volume of ``proppant'' (generally sand) is also injected in order to keep the fracture open. Traditional fracking is not effective in tight reservoirs. It is estimated that some 80\% of the producing wells in the United States have been subjected to traditional fracking \cite{Montgomery2010a}.

\begin{figure}
\centering
\begin{tabular}{c c}
 \includegraphics[width = 0.4\textwidth]{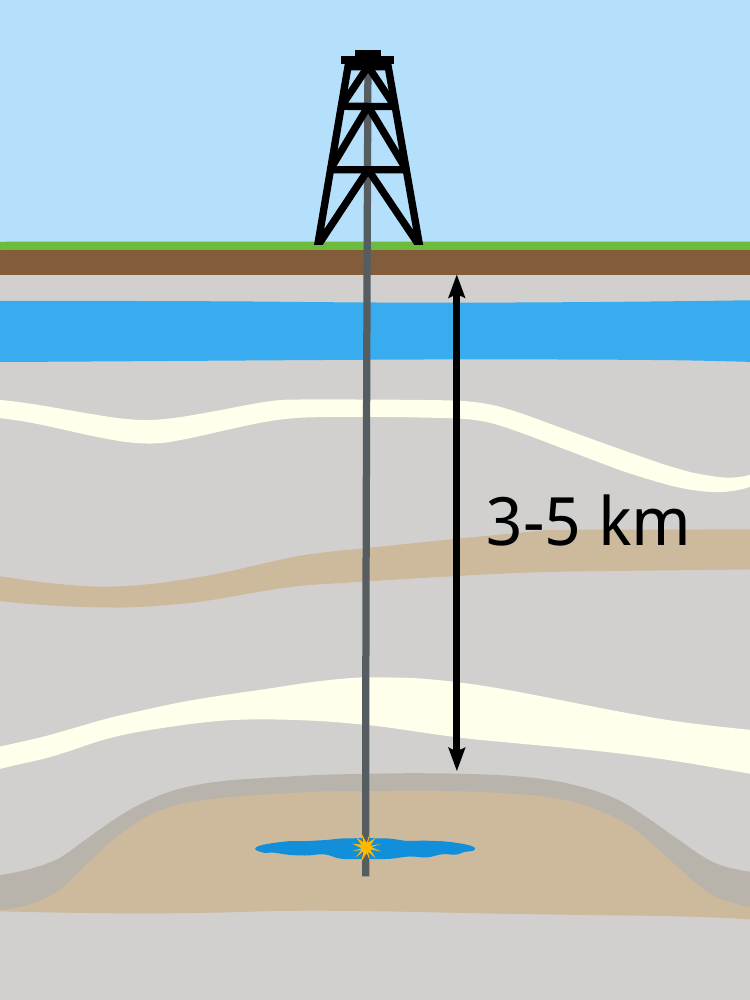} &
 \includegraphics[width = 0.4\textwidth]{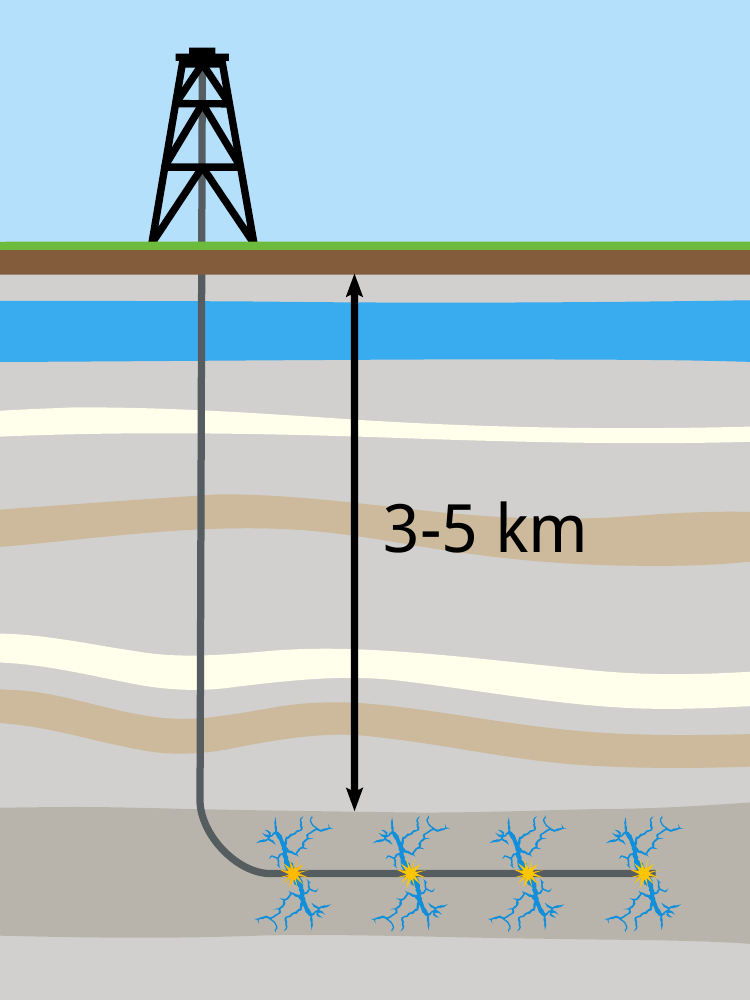} \\
 (a) & (b) \\
\end{tabular} 
 \caption{Schematic diagram of traditional and super fracking (a) Traditional fracking from a vertical production well. A high viscosity fluid is injected to create a single hydraulic fracture through which oil and/or gas migrate to the production well. (b) A sequence of 4 super fracking injections. Large volumes of a low viscosity fluid are injected to create a wide distribution of hydraulic fractures (damage). Oil and/or gas migrate through this network of fractures to the production well. Directional drilling produces a horizontal production well in the target strata allowing a sequence of super fracking injections to be carried out.}
 \label{fig:fracking_diagram}
\end{figure}

The second process is super (high volume) fracking, the primary advance that has made tight shale production possible. A typical super fracking injection uses $7.5 \times 10^3 \unit{m^3}$ to $10^4 \unit{m^3}$ of water, approximately 100 times as much water as in a traditional fracking injection. The development that has made high volume fracking possible is the use of ``slickwater'' as the injection fluid. ``Slickwater'' is a fluid in which the viscosity of the water is reduced by the addition of chemical additives, usually polyacrylamide \cite{Curtis2002}. This practice allows the injection of much larger volumes of water at the same injection pressure because of the reduction of viscosity and resultant resistance to flow of the water. Super-fracking is illustrated schematically in Fig. \ref{fig:fracking_diagram}.

The objective of super fracking is to create a large volume of damage in the reservoir, i.e. to create a widely distributed network of fractures through which oil and gas can migrate to the production well \cite{Busetti2012, Busetti2012b}. The production well is drilled vertically until it reaches the target strata including the production reservoir. Using directional drilling, the well is then extended horizontally into the target strata. The horizontal extension is typically several kilometers in length. It is desirable to target relatively deep, $3-5 \unit{km}$, reservoirs so that there is high lithostatic pressure to drive the fluid out. Plugs or ``packers'' are used to block off a section of the well, and explosives are used to perforate the well casing. Super-fracking injects ``slickwater'' through the blocked off perforation to create distributed hydrofractures as illustrated in Fig. \ref{fig:fracking_diagram}b. A sequence of super fracking injections are carried out as shown.

\begin{figure}
\centering
 \includegraphics[width=0.8\textwidth]{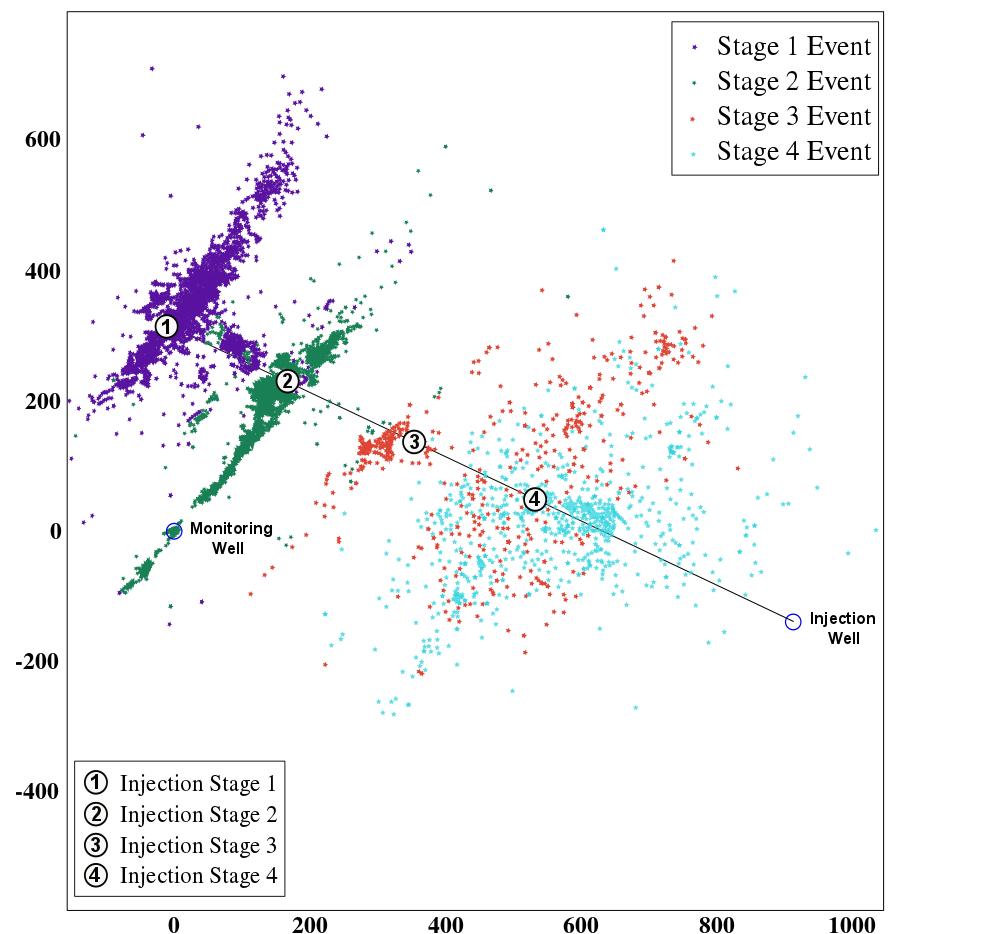}
 \caption{Map of the epicenter of small earthquakes associated with four super fracks of the Barnett shale in Texas \cite{Maxwell2011}. This microseismicity shows the distribution of fractures (damage) induced by the injected water. The axes are the distances (in meters) from the monitoring well.}
 \label{fig:microseismicity}
\end{figure}

Super-fracking creates a distribution of microseismicity that documents the complex fracture network that is being generated. In order to document the area that is being fractured, it is now standard practice to drill one or more vertical monitoring wells, with seismometers distributed along heir lengths. These seismometers can locate the microseismicity in real time and the results are used to control the rates of injection. Fig. \ref{fig:microseismicity} shows a typical example from the Barnett Shale in Texas \cite{Maxwell2011}. This map shows the epicenters from a four stage super fracking treatment as well as the locations of the vertical and horizontal components of the injection (production) well and the vertical monitoring well. The first and second injections produced relatively narrow clusters of seismicity while the the third and fourth injections produced much broader clusters indicating less localized fractures. The narrow clusters probably resulted from the orientation of the least principal compressional stresses parallel to the horizontal injection well. The wave forms of the larger events discriminate between deviatoric tensional failures or shear failures. These observations generally have large s-wave amplitudes relative to p-wave amplitudes indicating shear failures \cite{Maxwell2011, Rutledge2004}. The conclusion is that most of the events occur on preexisting healed natural fractures with a regional shear stress component generating shear displacement.

It is of interest to compare the role of super fracking in the extraction of gas from two relatively old tight black shales. We first consider the Barnett Shale in Texas, the site of the original development of super fracking injections of slickwater. Production rapidly accelerated in the early 2000's with the refinement of super fracking technology. So far some 8,000 wells have been drilled, about 90\% since the year 2000. Most of these wells are horizontal and have been subject to super fracking.

The Barnett Shale is a black shale of Late Mississippian age (323-340 Myr) located in the Fort Worth Basin. The organic carbon concentration in productive Barnett Shale ranges from less than 0.5\% to more than 6.0\%, with an average of 4.5\% by weight. Depths of production range from about $1.5$ to $2.5 \unit{km}$. The production formation has a maximum thickness of about $300 \unit{m}$, is relatively flat lying, and has only slight tectonic deformations.

Most natural hydraulic fractures in the Barnett Shale have been completely sealed by carbonate deposition \cite{Gale2007a} The bonding between the carbonate and shale is weak so that it is relatively easy for the super fracking injection to open the sealed fractures. There is strong evidence that open natural fractures prevent super fracking injections from creating distributed fractures. The injected slickwater leaks through the natural fractures without producing further damage.

We next consider the Antrim Shale in Michigan. The Antrim Shale is a tight black shale of Upper Devonian age ($354-370 
\unit{Myr}$) in the horizontally stratified Michigan Basin, Michigan. In terms of age, black shale deposition, and tectonic setting, the Antrim Shale is very similar to the Barnett Shale. Gas in the Antrim Shale is produced from some 9,000 wells, almost entirely traditional wells using traditional production techniques. This production utilizes open natrual fractures. Most fractures in the Antrim Shale are uncemented \cite{Curtis2002}. Because of this existing fracture permeability, super fracking does not appear to be effective. The low viscosity water used in a super fracking injection migrates through the preexisting, open natural fractures without effectively increasing the permeability. The inability of super fracking to increase production can explain the decline in production of gas from the Antrim Shale at the same time that gas production from the Barnett Shale was rapidly increasing.

\section*{Our model}
\begin{figure}
\centering
\includegraphics[scale=1]{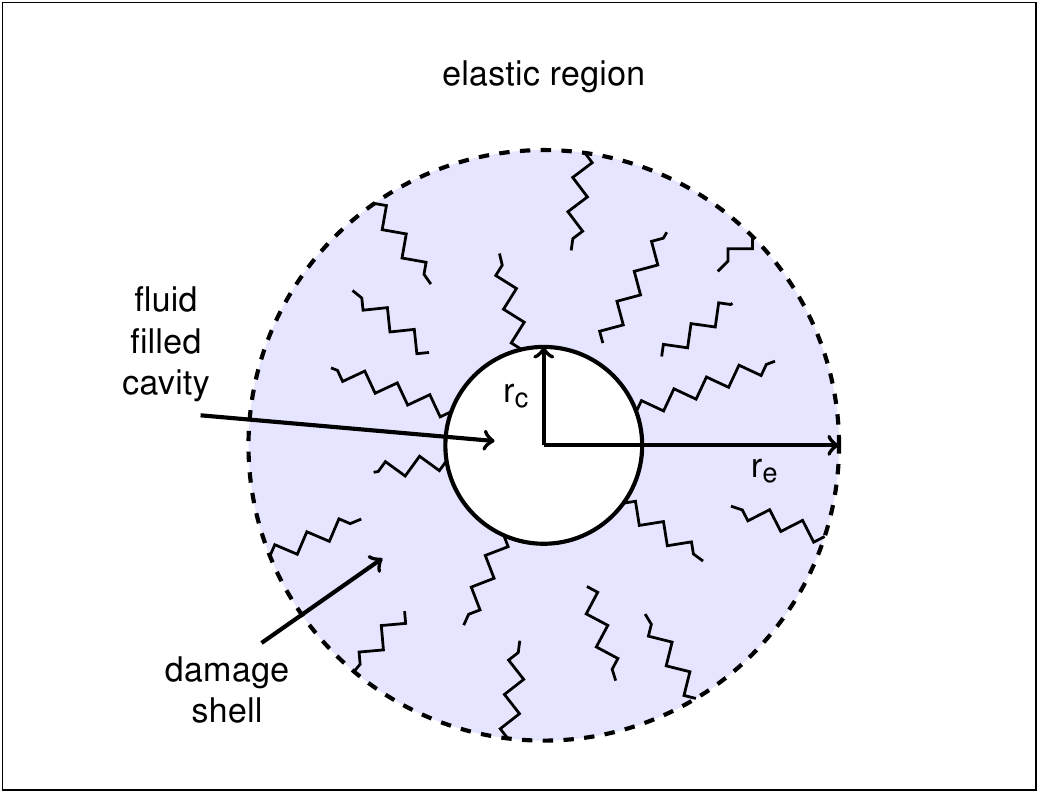}
\caption{Illustration of our spherically symmetric  problem for fracking. Fluid from the fluid filled cavity (radius $r_c$) infiltrates the damaged shell with outer radius $r_{e}$ inducing radial fractures. The damaged rock is at the fluid pressure $p$. For $r>r_e$ the rock exhibits uniform elastic behavior.}
\label{fig:SphericalCavity}
\end{figure}

In order to better understand the fundamental processes associated with super fracking we will consider a relatively simple spherically symmetric problem. This problem is illustrated in Figure \ref{fig:SphericalCavity}. Fluid is injected from a spherical fluid filled cavity with radius $r_c$. Initially the cavity has a radius $r_{c0}$ and it is embedded in a uniform infinite elastic medium. We assume the medium and fluid are initially at a uniform lithostatic pressure $p_{L} = \rho g h$ where $h$ is the depth of the cavity. This assumption assumes that the radius influenced by a high fluid pressure is small compared with $h$ (generally a good approximation). In the solution given below, all pressures, stresses and strains are given as variations from the uniform background conditions, pressure $p_{L}$. In order to initiate fluid fracturing, the fluid pressure $p$ is increased. At relatively low fluid pressures the surrounding rock deforms elastically. The elastic solution for the spherically symmetric stress and strain fields resulting from a pressurized (pressure $p$) fluid filled cavity is given by \cite{Galanov2008}
\begin{eqnarray}
\sigma_{r} = p \left(\frac{r_e}{r}\right)^3, && \sigma_{h} = -\frac{p}{2}\left(\frac{r_e}{r}\right)^3
\label{eq:stress}
\end{eqnarray}
\begin{eqnarray}
\epsilon_{r} = \frac{(1+\nu)}{E} p \left(\frac{r_e}{r}\right)^3, && \epsilon_{h} = -\frac{(1+\nu)}{2 E} p\left(\frac{r_e}{r}\right)^3
\label{eq:strain}
\end{eqnarray}
where $r_{e}$ is the inner radius of the elastic region, $\sigma_r$ and $\epsilon_r$ are the compressional radial components of stress and strain, and $\sigma_{h} =\sigma_{\theta} =\sigma_{\phi}$ and $\epsilon_{h} =\epsilon_{\theta} =\epsilon_{\phi}$ are the compressional hoop (azimuthal) components of the spherically symmetric stress and strain fields.

The fluid pressure is increased to the value $p_d$ at which damage (micro-cracking) occurs. The hoop stress becomes tensional when $\sigma_h=-p_{L}$. We assume that damage occurs when the maximum elastic hoop stress $\sigma_{hd}$ is a specified fraction $f$ of the lithostatic pressure $p_{L}$
\begin{equation}
\sigma_{hd}=-f p_L
\label{eq:damage_condition}
\end{equation}
with $0<f<1$. This condition allows shear fractures to develop in the damage zone even though both $\sigma_r$ and $\sigma_h$ are compressional.

As more fluid is injected, a spherical shell of damaged rock is created with an outer radius $r_e$ (the inner radius of the elastic region). We assume that the damaged shell of rock has connected radial fractures that allow fluid to penetrate and eliminate differential hoop stresses in the damaged shell. We further assume that the injection rate of fluid is sufficiently small that the pressure drop associated with the fluid flow through the damaged shell can be neglected. Without hoop stresses, the stress field in the damaged shell is isotropic and equal to the fluid pressure $p_d$. From Eqs. \eqref{eq:stress} and \eqref{eq:damage_condition} the components of stress and strain at the inner boundary of the elastic region ($r=r_e$) are given by
\begin{eqnarray}
\sigma_{r}\left(r_e\right) = p_d, && \sigma_{h}\left(r_e\right) = -\frac{p_d}{2}
\label{eq:stress_boundary}
\end{eqnarray}
\begin{eqnarray}
\epsilon_{r}\left(r_e\right) = \frac{(1+\nu)}{E} p_d, && \epsilon_{h}\left(r_e\right) = -\frac{(1+\nu)}{2 E} p_d
\label{eq:strain_boundary}
\end{eqnarray}

From Eqs. \eqref{eq:damage_condition} and \eqref{eq:stress_boundary} the fluid pressure $p_d$ in excess of the lithostatic pressure required to generate damage is given by
\begin{equation}
p_d = 2 f p_L
\label{eq:damageP}
\end{equation}
As stated above, this value is independent of the thickness of the damaged shell. In our solution, we specify the inner radius of the elastic region $r_e$ and will determine the required volume of injected fluid necessary to generate the required porosity.

The increase in the volume of the damaged shell $\Delta V_1$ due to the increase in the radius of the elastic region from $r_{e0}$ to $r_e$ is given by 
\begin{equation}
 \Delta V_1 = 4 \pi r_{e0}^2\left(r_e-r_{e0}\right)
 \label{eq:volume_change_1}
\end{equation}
assuming the condition for linear elasticity that $\frac{\left(r_e-r_{e0}\right)}{r_{e0}} \ll 1$. The change in radius is related to the elastic hoop (azimuthal) strain at $r=r_{e0}$ by
\begin{equation}
 r_e - r_{e0} = -r_{e0} \epsilon_h \left(r_{e0}\right)
 \label{eq:radial_strain_1}
\end{equation}
Substitution of Eqs. \eqref{eq:strain_boundary} and \eqref{eq:damageP} into Eq. \eqref{eq:radial_strain_1} gives
\begin{equation}
 r_e-r_{e0} = \frac{r_{e0} \left(1 + \nu \right) f p_L}{E}
 \label{eq:radial_strain_2}
\end{equation}
An substitution of Eq. \eqref{eq:radial_strain_2} into \eqref{eq:volume_change_1} gives the increase in the volume of the damaged shell due to the decrease in the volume of the elastic region
\begin{equation}
 \Delta V_1 = \frac{4 \pi r_{e0}^3\left(1+\nu\right) f p_L}{E}
 \label{eq:volume_change_2}
\end{equation}
As discussed above we assume that the damaged rock has a uniform increase in pressure $p_d$. The compressibility of the damaged rock is given by
\begin{equation}
 \beta = \frac{3 \left(1-2\nu\right)}{E}
 \label{eq:compressibility}
\end{equation}
The increase in the volume of the damaged shell $\Delta V_2$ due to the decrease in the volume of the damaged rock is given by
\begin{equation}
 \Delta V_2 = \frac{4}{3} \pi r_{e0}^3 \beta p_d
 \label{eq:volume_change_3}
\end{equation}
In writing this relation we have neglected the volume of the spherical fluid filled cavity i.e. $r_c^3 \ll r_e^3$. Substitution of Eqs. \eqref{eq:damageP} and \eqref{eq:compressibility} gives
\begin{equation}
 \Delta V_2 = \frac{8 \pi r_{e0}^3 \left(1- 2\nu\right) f p_L}{E} 
 \label{eq:volume_change_4}
\end{equation}

The required volume of fracking fluid $\Delta V_f$ needed to generate a damage region of radius $r_e$ is given by
\begin{equation}
 \Delta V_f = \Delta V_1 + \Delta V_2 = \frac{12 \pi r_{e0}^3 \left(1- \nu\right) f p_L}{E} 
 \label{eq:fluid_volume}
\end{equation}
The volume $\Delta V_f$ is the volume generated by damage (fractures) in the spherical shell.

We relate the damage volume $\Delta V_f$ generated by the fluid pressure $p_d$ to the elastic volume change that is generated by the fluid pressure by introducing the damage variable $\alpha$ defined by
\begin{equation}
 \frac{\Delta V_f}{\frac{4}{3}\pi r_{e0}^3} = \left(\frac{\alpha}{1-\alpha}\right) \beta p_d 
 \label{eq:damage_definition}
\end{equation}
When $\alpha = 0$ we have $\Delta V_f = 0$ and there is no damage. In the limit $\alpha \to 1$ we have $\Delta V_f \to \infty$ and the damaged rock disintegrates. Substitution of Eqs. \eqref{eq:damageP} and \eqref{eq:compressibility} into Eq. \eqref{eq:fluid_volume} gives
\begin{equation}
 \frac{\Delta V_f}{\frac{4}{3}\pi r_{e0}^3} = \frac{3}{2} \left(\frac{1-\nu}{1-2 \nu}\right) \beta p_d 
 \label{eq:damage_expression}
\end{equation}
Comparison of Eqs. \eqref{eq:damage_definition} and \eqref{eq:damage_expression} gives
\begin{equation}
 \alpha = \frac{3 \left(1-\nu\right)}{\left(5-7 \nu\right)}
 \label{eq:damage}
\end{equation}
For shale we take $\nu=0.17$ and find that $\alpha=0.654$. The damage variable is only a function of the Poisson ration $\nu$ and does not depend on the thickness of the damaged shell.

We next estimate the permeability of the damaged shell. We approximate the structure of the damaged region to be a cubic matrix with dimensions $b$. The walls of each cube are channels of uniform width $\delta$. These channels approximate the joint sets generated by natural fracking. The porosity $\phi$ of this geometry is given by
\begin{equation}
 \phi = \frac{\Delta V_f}{\frac{4}{3}\pi r_{e0}^3} = 3 \frac{\delta}{b}
 \label{eq:porosity_equation}
\end{equation}
We assume that the pressure gradient is perpendicular to one face of the cube. In this case, the permeability $k$ is given by
\begin{equation}
 k = \frac{1}{6} \frac{\delta^3}{b}
 \label{eq:permeability_equation}
\end{equation}
We next consider a specific example related to super fracking.

\section*{An example}
\begin{figure}
\centering
\includegraphics[scale=1]{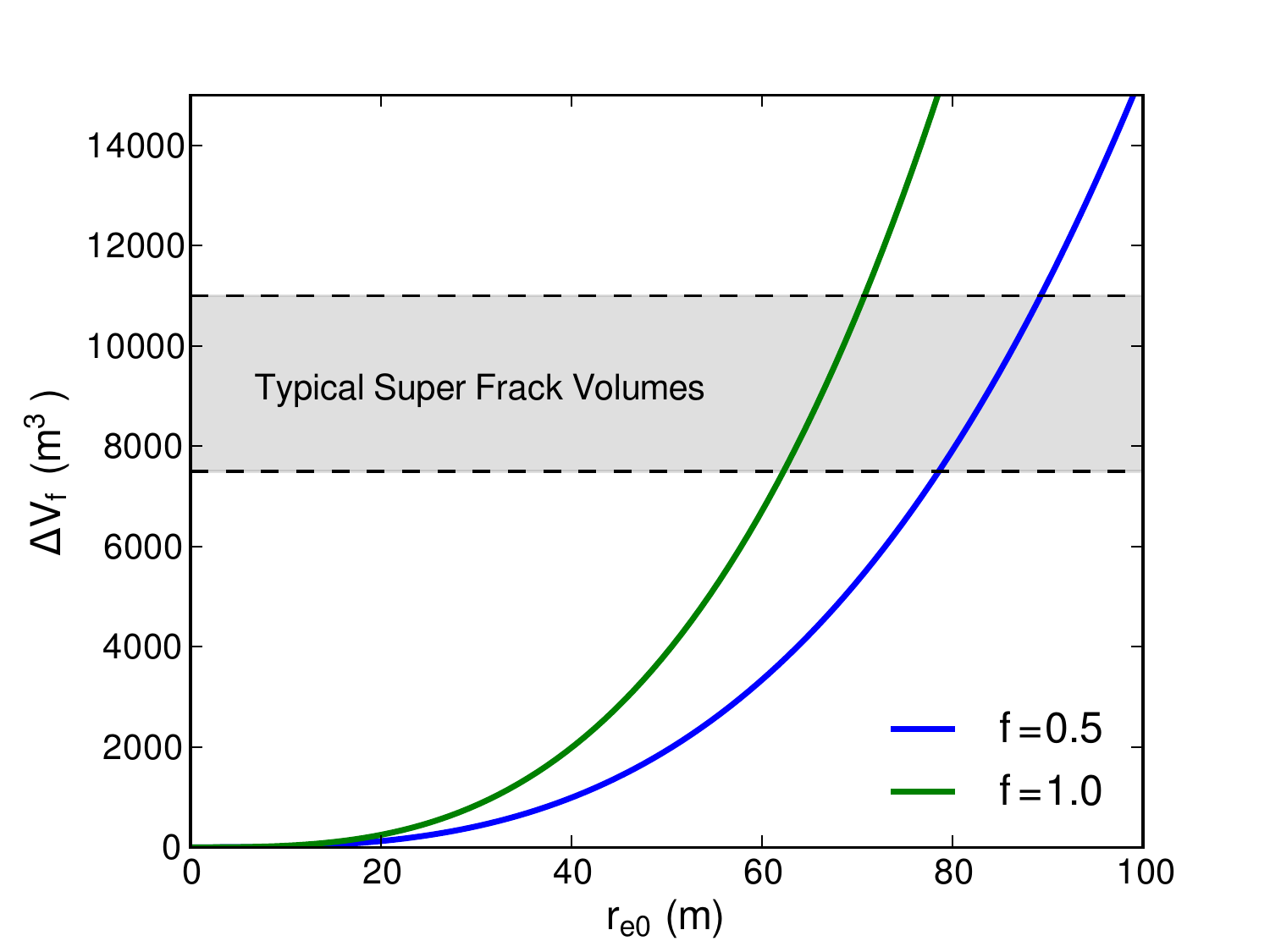}
\caption{Required fluid volume $\Delta V_f$ as a function of damage shell radius $r_{e0}$ for two values of $f$  with $\rho = 2,620\unit{kgm^{-3}}$, $E = 65 \unit{GPa}$, $\nu = 0.17$, and $h=2.5\unit{km}$. Also shown is the typical range of water volumes used in super fracks.}
\label{fig:fluid_volume}
\end{figure}

In order to test our model we will consider a specific example. A typical well diameter is 0.15\unit{m} and we will take this to be the radius of our initial fluid filled cavity $r_{c0}=0.15\unit{m}$. For the properties of the oil (gas) shale, we take the density $\rho = 2,620\unit{kg \cdot m^{-3}}$, Young's modulus $E = 65 \unit{GPa}$, and Poisson's ratio $\nu = 0.17$ \cite{Engelder1990}. As a typical tight shale reservoir we will take the Barnett shale in Texas that was previously discussed. Production depths are in the range 2-3 \unit{km} and shale layer thicknesses are in the range 10 to $300\unit{m}$. For our example we will take $h=2.5\unit{km}$. The lithostatic pressure in the reservoir is $p_L = \rho g h = 64 \unit{MPa}$. We first give the volume of water required to frack a spherical shell of radius $r_e$ using Eq. \eqref{eq:fluid_volume}. The dependence of the fluid volume $\Delta V_f$ on the damage radius $r_{e0}$ is given in Fig. \ref{fig:fluid_volume} taking $f=1.0$ and $0.5$. As discussed in the introduction the typical range of water values used in super fracks is $7.5\times10^3\unit{m^3}$ to $11.0\times10^3\unit{m^3}$. This range is also shown in Figure \ref{fig:fluid_volume}. Our results indicate that this volume of water would produce a sphere of damaged shale with a radius in the range $60$ to $90\unit{m}$.

We now estimate the damage volume for the State 1 frack illustrated in Fig. \ref{fig:fluid_volume}. Clearly the horizontal distribution of damage associated with microseismicity is not circular, but we can approximate the area to be $300 \times 75 = 2.25 \times 10^4 \unit{m^3}$. We also approximate the vertical height of damage to be $75 \unit{m}$ for a total damage volume of $1.69 \times 10^6 \unit{m^3}$. The radius of a sphere with this volume is $r= 74 \unit{m}$. This is within the range of values given in Fig. \ref{fig:fluid_volume}. Our comparison is certainly approximate both in terms of the shape of the damage region and the uniformity of damage in the region. But we suggest that the agreement indicates our basic approach to calculating the porosity (water) volume is approximately correct.

From Eqs. \eqref{eq:fluid_volume} and \eqref{eq:porosity_equation} the porosity of the damaged region is
\begin{equation}
 \phi = \frac{9\left(1-\nu\right)}{E} f p_L
 \label{eq:porosity}
\end{equation}
Taking values given above we find that the porosity with $f=1$ is $\phi = 7.4 \times 10^{-3}$ and with $f=0.5$ $\phi=3.7 \times 10^{-3}$. With our assumptions these values are independent of the volumes of the damaged rock. For our assumed grid of cubic fractures, the open fracture width from Eq. \eqref{eq:porosity_equation} is
\begin{equation}
 \delta = \frac{b \phi}{3}
 \label{eq:fracture_width}
\end{equation}

Taking fractures spacings $b=0.1 \unit{m}$ and $b=1.0 \unit{m}$ we find fracture thickness of $\delta=0.25 \unit{mm}$ and $2.5 \unit{mm}$ for $f=1$ and $\delta=0.125 \unit{m}$ and $1.25 \unit{mm}$ for $f=0.5$. For our model the permeability is given by Eq. \eqref{eq:permeability_equation}. For fracture spacings $b=0.1 \unit{m}$ and $b=1.0 \unit{m}$ we find permeabilities $k=2.6 \times 10^{-11} \unit{m^2}$ and $k=2.6 \times 10^{-9} \unit{m^2}$ for $f=1$ and $k=3.25 \times 10^{-12} \unit{m^2}$ and $k=3.25 \times 10^{-10} \unit{m^2}$ for $f=0.5$. Typical sandstone reservoirs have permeabilities $k$ in the range $10^{-12}$ to $10^{-14} \unit{m^2}$ so our model produces ample permeabilities for the extraction of oil and gas. For comparison the permeabilities of tight shales are in the range $k=10^{-19}$ to $10^{-21} \unit{m^2}$.

\section*{Discussion}
Large volume ``super'' fracking is effective in extracting oil and/or gas from tight shale reservoirs. In a super frack some $10^4 \unit{m^3}$ of water is injected from a well. It is important that the water includes additives that reduce its viscosity creating ``slick" water. Studies of the resulting microseismicity indicate that the injected water migrates along preexisting fractures resulting in small shear seismic displacements. The objective of a super frack is to create a network of fractures with significant permeability to allow the oil and/or gas to migrate to the well when the injected water is removed.

The purpose of this paper is to present a relatively simple model for super fracking. We consider a spherically symmetric problem with fluid injection from a small spherical cavity (modeling the injection from a perforated well bore). The high pressure of the injected water generates hoop stresses that reactivate preexisting natural fractures in the tight shale. These fractures migrate outward as water is added creating a spherical shell of damaged rock. We neglect the pressure drop associated with this flow so that the radius of the damaged shell is controlled by the volume of water injected.

In order to test our model we have considered a specific example typical of a tight shale reservoir. The fluid volumes predicted by our model are compared with actual super fracking volumes in Figure \ref{fig:fluid_volume}. It is seen that there is quite good agreement.

We recognize that our model is highly idealized. However, we believe it provides a basis for understanding the role of super fracking in generating permeability. A number of our assumptions are certainly subject to uncertainties. Examples include:

1) We assume that the fracking induced permeability is associated with the reactivation of the preexisting natural fractures. Another possible mechanism is that the super fracking breaks up (comminution) the shale. This would induce permeability on a much smaller scale but would also require large amounts of energy. We believe that the observed microseismicity favors our model. Also, the role of natural fractures in the extraction of shale oil and gas has been discussed in considerable detail by Engelder et al. \cite{Engelder2009}.

2) We assume that the initial stress state is isotropic. This is clearly not the case and the stress variability will focus the individual fractures. In general, fractures will open perpendicular to the minimum principal stress. In shale reservoirs, the maximum principal stress is generally vertical. Horizontal wells are generally drilled in the direction of the minimum principal stress so fractures will tend to be oriented perpendicular to the horizontal well. Observations of super fracking induced microseismicity \cite{Maxwell2011} indicate considerable deviations in fracture orientation. A future extension of our present study could include a non-isotropic stress field.

\section*{Funding}
The research of JQN and JBR has been supported by a grant from the US Department of Energy to the University of California, Davis, \#DE-FG02-04ER15568

\end{document}